\documentclass{epl}

\usepackage{graphicx}

\title{Growing network with $j$-redirection}

\author{R. Lambiotte\inst{1} and M. Ausloos\inst{1}}

\institute{ \inst{1}GRAPES, Universit\'e de Li\`ege, B5 Sart-Tilman, B-4000 Li\`ege, Belgium}


\begin{document}

\maketitle

\begin{abstract}
A model for growing information networks is introduced where nodes receive new links through $j$-redirection, i.e. the probability for a node to receive a link depends on the number of paths of length $j$ arriving at this node. In detail, when a new node enters the network, it  either connects to a randomly selected node, or to the $j$-ancestor of this selected node. The $j$-ancestor is found by following $j$ links from the randomly selected node. The system is shown to undergo a transition to a  phase where condensates develop. We also find analytical predictions for the height statistics and show numerically the non-trivial behaviour of the degree distribution.
\end{abstract}
\date{\today}

\section{Motivation}

It is well-known that  large networked information systems (e.g. citation networks or the Web) are explored by following the links between items \cite{sidneyGoogle}. This process is at the heart of common search engines like Google, and is based on the  empirical observation that
an individual surfing the Web will typically follow of the order of 6 hyperlinks before switching to an unrelated site \cite{google}. Practically, search engines mimic this behaviour  by sending  "random walkers"  who, part of the time, follow links between websites, and otherwise jump to a randomly selected website in the network.
The average number of walkers at a given node is the measure of the importance of the node in the network (e.g. the Google Rank number).
In view of this search mechanisms, one expects that nodes with a higher density of walkers are visited more often, and should therefore receive more links from newly introduced nodes.
This feed-back mechanism leads to an increase of  the selected node degree, in a manner that may naively remind one of preferential attachment \cite{bar}, as well as its density of walkers, thereby increasing the probability of the selected node to be chosen in the future, etc...

In its most basic form, a growing network with redirection is defined as follows: a node enters the system, first connects to a target node (chosen randomly in the whole network) and then, with some probability $p$, redirects its link to the ancestor of the target node. This model is well-known \cite{krapi} to lead to linear preferential attachment in the network, and to reproduce the formation of fat tail degree distributions $k^{-\nu}$, with $\nu=1+1/p$. However, more realistic situations where the entering node follows $j \neq 1$ links before connecting to a node (see Fig.1) have not been considered yet. From now on, we call this recursive exploration of the network  $j$-redirection. In the following, we will mainly focus on the $2$-redirection case and restrict the scope to networks where nodes have only one outgoing link. We will show how this slight generalization leads to much more complicated situations than in the case $j=1$, such as the formation of condensates in the network.

\section{Basic model}

\begin{figure}
\hspace{1.9cm}
\includegraphics[width=4.0in]{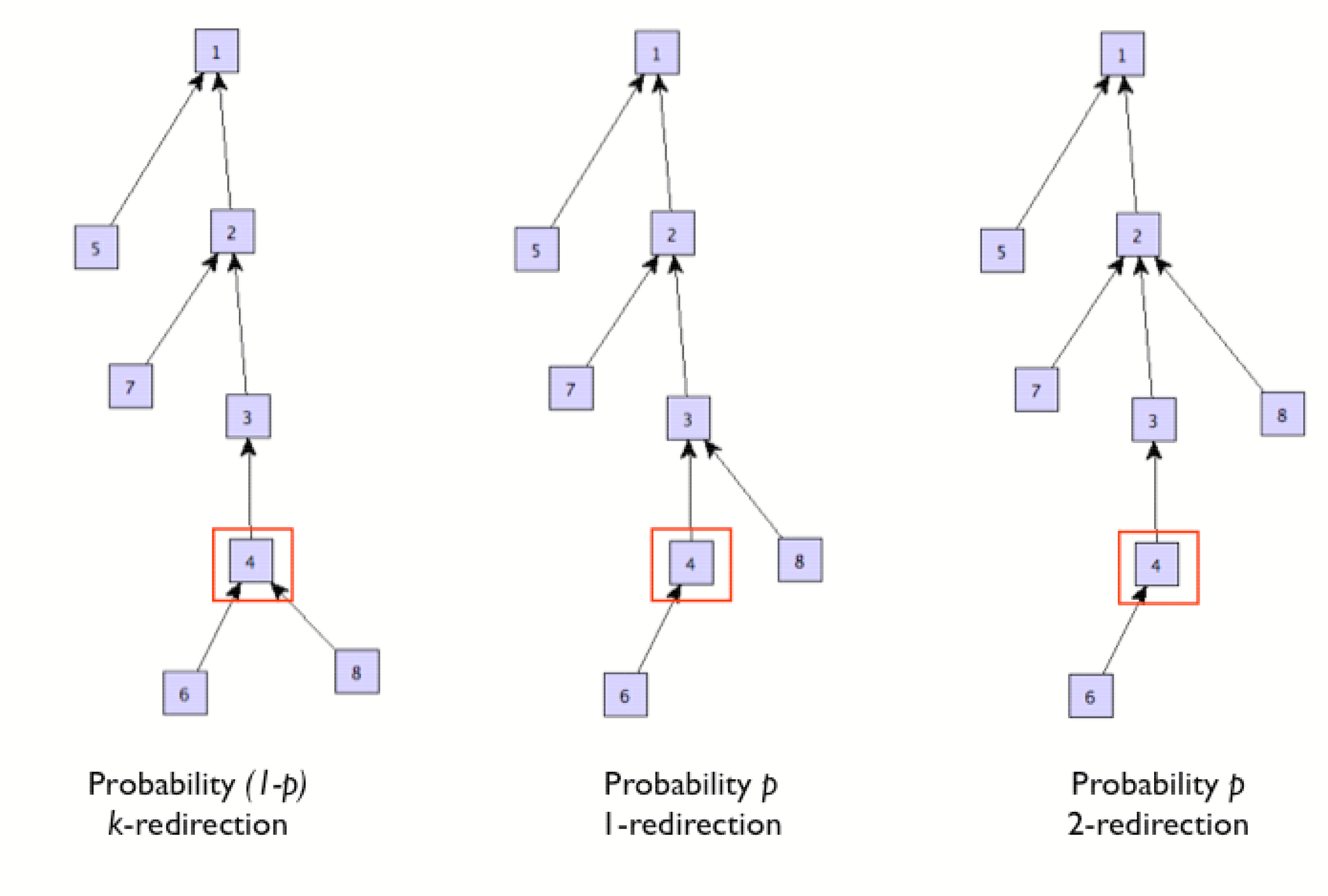}

\caption{Sketch of a time step of the model with 1-redirection or 2-redirection. The system is first composed of 7 nodes. A new node, labelled with 8, enters the network and randomly selects the marked node (node 4). If 1-redirection takes place, the entering node connects to the father of the marked node with probability $p$. If 2-redirection takes place, the entering node connects to the grand-father of the marked node with probability $p$. Otherwise, the entering node connects to the marked node.}
\label{kRedir}      
\end{figure}

Let us first study the simplest version of the model where entering nodes explore the network with $1$-redirection. Initially ($t=0$), the network is composed of one node, the seed, and each time step $t$, a new node enters the network.  Consequently, the total number of nodes is equal to $N=1+t$, and the number of links is $L=t$. We will focus on the height distribution,    
 the height of 
a node \cite{bennaim} being defined to be the minimum number of links 
to the seed. The probability that a node at the depth $g$ in the directed network receives the link is:
\begin{eqnarray}
P_g \sim (1-p) N_g + p N_{g+1}, 
\end{eqnarray}
except for the seed $g=0$:
\begin{eqnarray}
P_0 \sim  N_0 + p N_1, 
\end{eqnarray}
where  $N_g$ is the average number of nodes at depth $g$.
 The normalisation follows:
\begin{eqnarray}
 N_0 + p N_1  + \sum_{i=1}^{\infty} [(1-p) N_i + p N_{i+1} ]
= N.
\end{eqnarray}
Putting the above pieces together, it is straightforward to show that the rate equation for $N_g$ reads in the  continous time limit:
\begin{eqnarray}
\label{basic}
\partial_t N_{1;t} &=&  \frac{1}{N} (N_0 + p N_1) \cr
\partial_t N_{g;t} &=& \frac{1}{N} [(1-p) N_{g-1} + p N_{g} ] .
\end{eqnarray}
As a first level of description, we derive an equation for the average total height $G=\sum_{g=0}^{\infty} g N_g$ from Eq.{\ref{basic}}, that reads  in the long time limit $t\gg1$:
\begin{eqnarray}
\partial_t G &=&  [   (1- p)  +   \frac{G}{t} ] .
\end{eqnarray}
This equation  leads to the asymptotic behaviour $G \sim (1-p) t \ln t$, from which one recovers the behaviour $G \sim t \ln t$ taking place when the model is purely random (no redirection).
Consequently, the redirecting process slows down the growth of the network.  This is expected as redirection favours the connection to nodes closer to the seed. In the limiting case $p=1$, where the process is easily shown to lead to a star network (i.e. all the nodes are connected to the seed), one finds $G \sim t$. 

\begin{figure}
\includegraphics[width=1.9in]{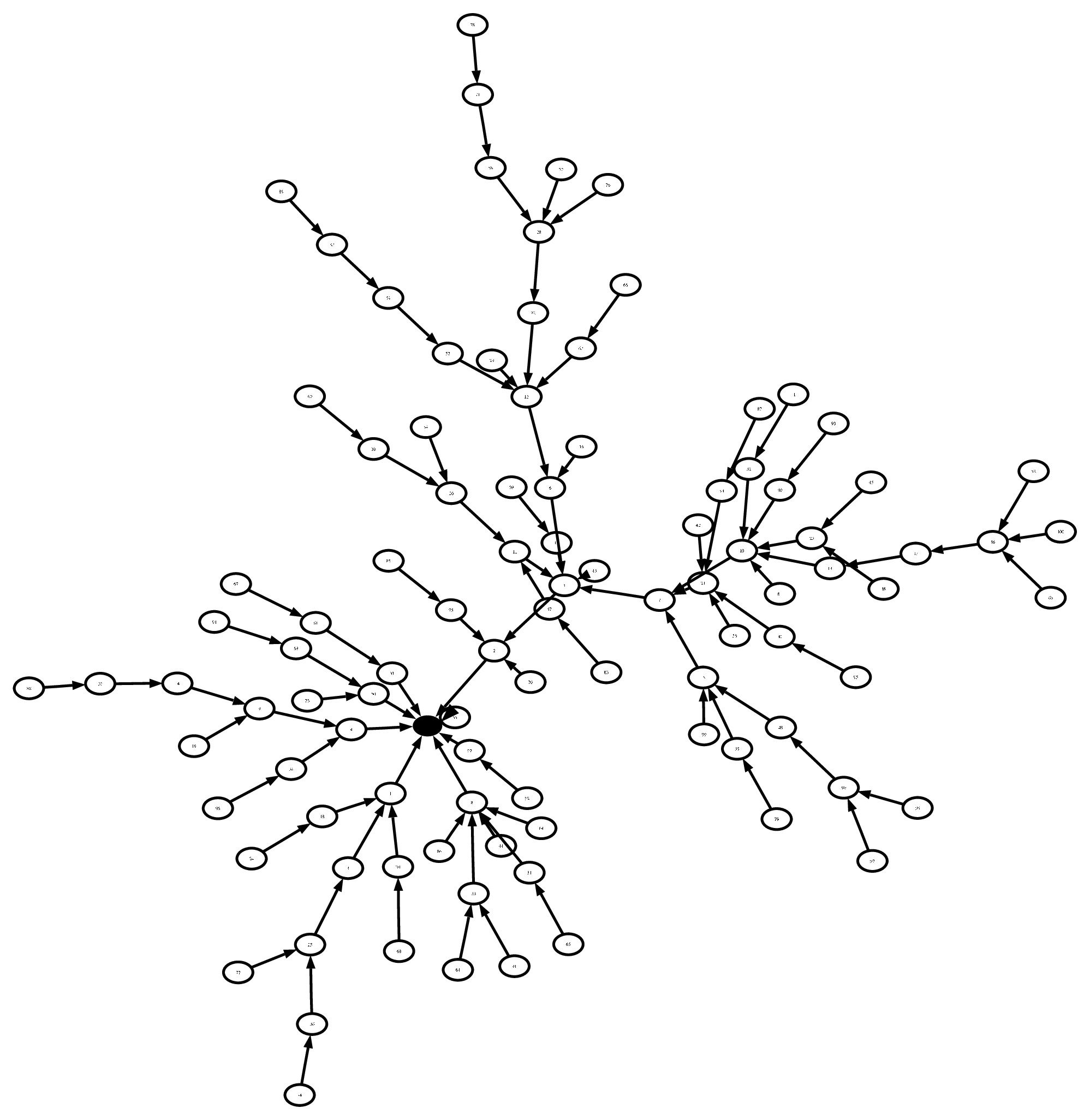}
\includegraphics[width=1.6in]{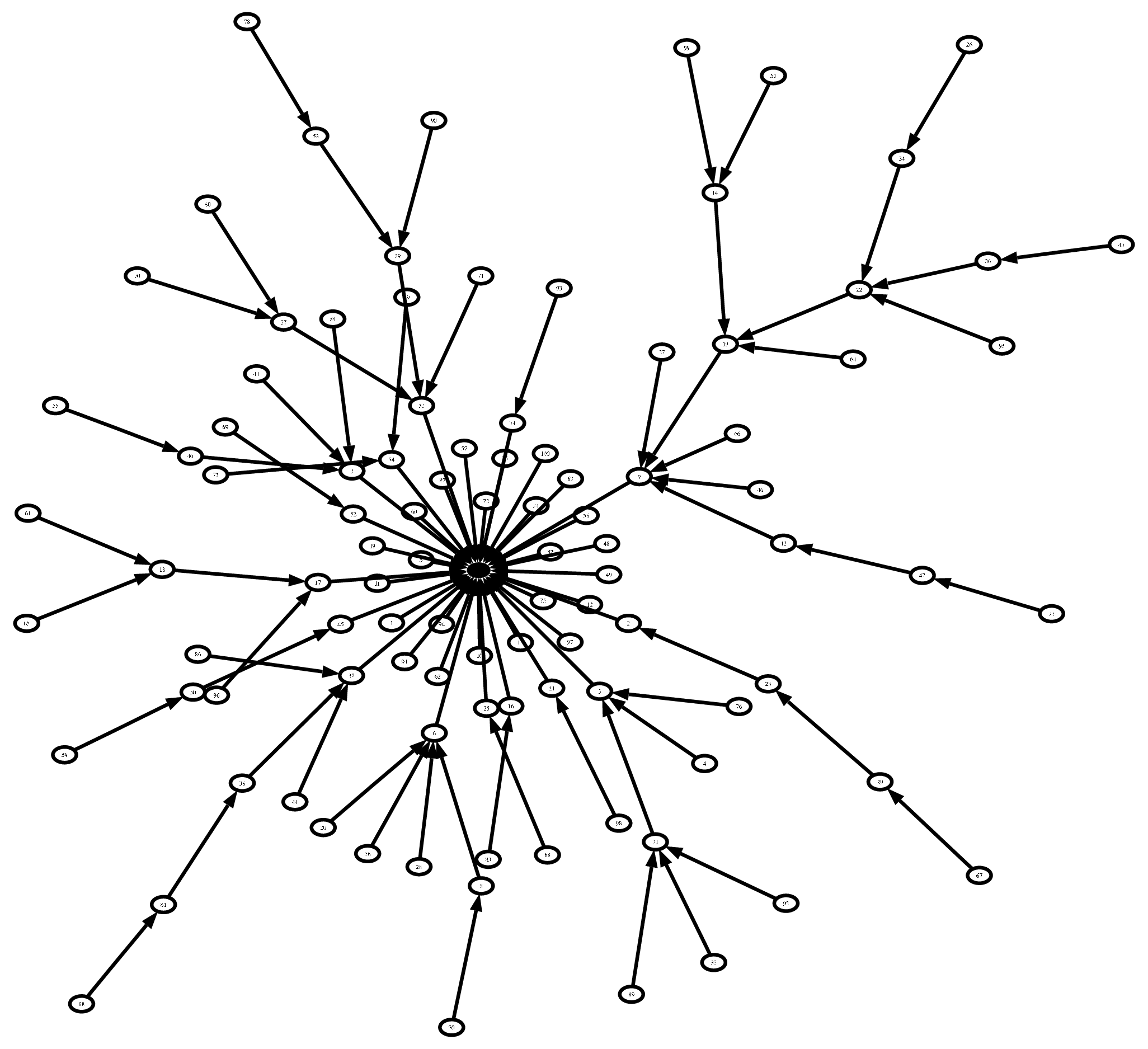}
\includegraphics[width=1.5in]{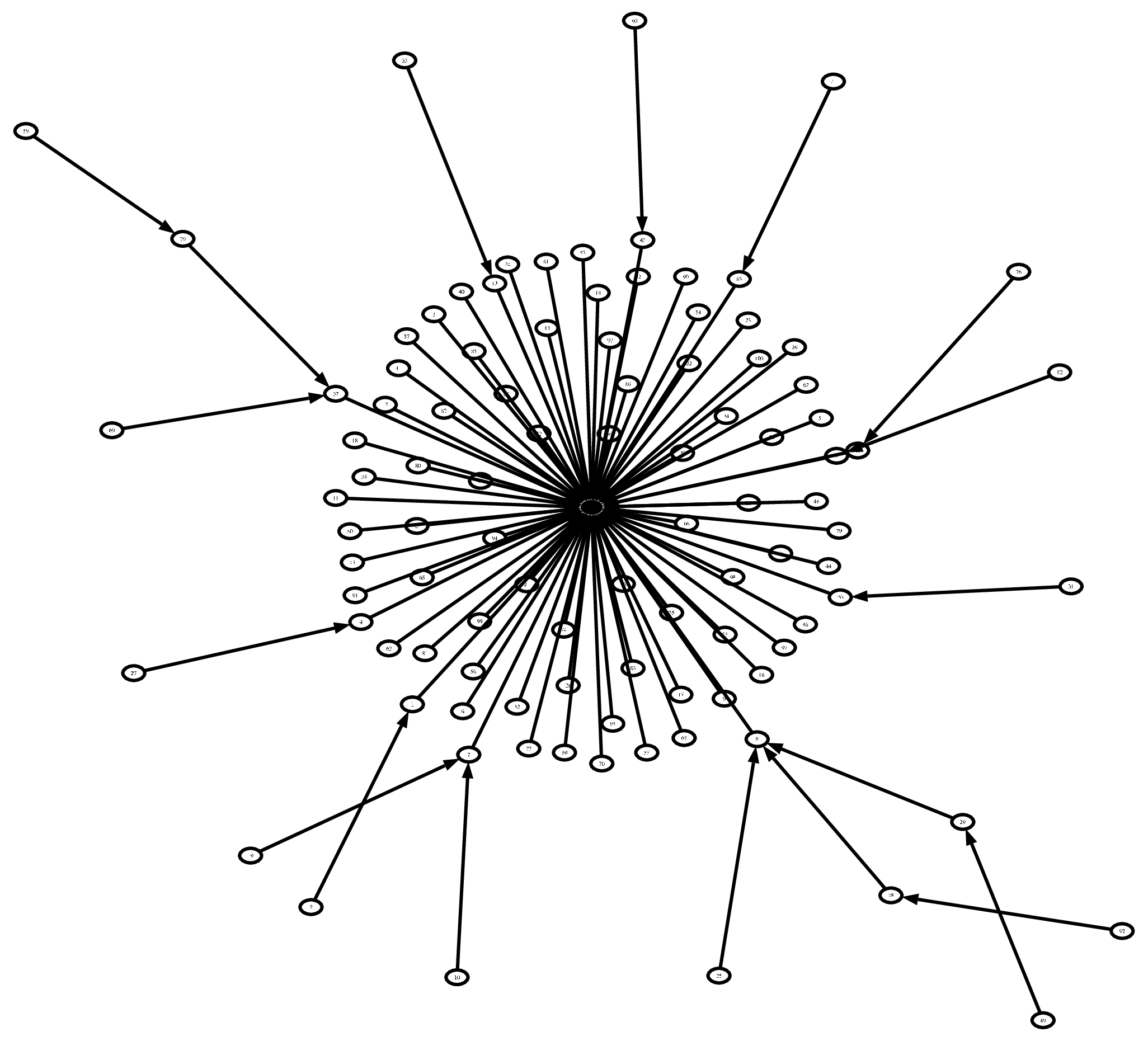}

\caption{Typical realizations of the 2-redirection model after 100 time steps for $p=0.0$, $p=0.4$ and $p=0.8$ (left to right).}
\label{typical}      
\end{figure}

\section{Condensation in the 2-redirection model}
Let us now focus on the more challenging case when the network is explored with $2$-redirection  (see Fig.2).
 The generalization to any value of $j>1$ is straightforward and will be briefly discussed at the end of this section.
The probability that a node at the depth $g$ in the directed network receives the link is:
\begin{eqnarray}
P_g \sim (1-p) N_g + p N_{g+2}, 
\end{eqnarray}
except for the seed, where:
\begin{eqnarray}
P_0 \sim  N_0 + p N_1 + p N_2,
\end{eqnarray}
and where the normalization follows:
\begin{eqnarray}
 N_0 + p N_1 + p N_2 + \sum_{i=1}^{\infty} [(1-p) N_i + p N_{i+2} ]
= N.
\end{eqnarray}
The rate equation for $N_g$ and the equation for the average $G$ are respectively:
\begin{eqnarray}
\label{cond}
\partial_t N_{1;t} &=&  \frac{1}{N} (N_0 + p N_1 + p N_2) \cr
\partial_t N_{g;t} &=& \frac{1}{N} [(1-p) N_{g-1} + p N_{g+1} ], 
\end{eqnarray}
and
\begin{eqnarray}
\label{eqG}
\partial_t G &=&  [  p n_1  + (1-2 p)  +   \frac{G}{t} ],
\end{eqnarray}
where $n_g\equiv N_g/N$ is the proportion of nodes at height $g$.
There are obviously two possible cases. i) If $n_1$ is vanishingly small in the long time limit, Eq.\ref{eqG} simplifies into
\begin{eqnarray}
\partial_t G =   (1-2 p)  +   \frac{G}{t},
\end{eqnarray}
whose solution is $G \sim (1-2p) t \ln t$. This solution suggests that a qualitative change occurs around $p_c=1/2$.  ii) If $n_1$ does not vanish in the long time limit, this term has to be taken into account. Let us stress that a finite value of $n_1$ implies the formation of a condensate in the network, i.e. the seed attracts a non-vanishing fraction of the links in the network \cite{con1,con2,con3,con4}. 

Let us evaluate the values of $p$ for which such a condensate exists and the corresponding value of $n_1$. To do so, one needs to find stationary solutions to the equations for  $n_g$:
\begin{eqnarray}
(1+t) \partial_t n_{1} &=&  (n_0 + p n_1 + p n_2) - n_1\cr
(1+t) \partial_t n_{g} &=&  [(1-p) n_{g-1} + p n_{g+1} ] - n_g.
\end{eqnarray}
The stationary solution are found by recurrence and by using the fact that $n_0$ is negligible in the long time limit. Indeed, $N_0$ is (and remains) equal to $1$ by construction, so that $N_0/N = 1/N \rightarrow 0$. It is straighforward to show that the stationary solution is in general:
\begin{equation}
\label{eigenvector}
n_{g} =  \frac{1}{C}\left(\frac{ 1-p}{p}\right)^{g-1},
\end{equation}
whose normalisation constant is $C= \sum_{g=1}^{\infty} (\frac{ 1-p}{p})^{g-1}$.
Consequently, the system reaches a stationary solution when $p>1/2$ and $\frac{ 1-p}{p}<1$, so that $C=\frac{p}{2p-1}$.  Otherwise, the probability normalisation is not  satisfied.

By inserting the above solution $n_1=  \frac{2p-1}{p}$ into Eq.\ref{eqG}, one arrives at the trivial evolution equation
\begin{eqnarray}
\partial_t G &=&  \frac{G}{t}, 
\end{eqnarray}
so that the average height $G/t$ asymptotically goes to a constant, in agreement with the observed formation of condensates. Before focusing on the regime $p<1/2$, let us stress that 
the existence of non-vanishing stationary values of $n_g$ is not possible in the 1-redirection model. In contrast, the formation of condensates takes place for any other $j$-redirection  $j>1$. This result is straightforward after generalizing Eq.\ref{eqG} into:
\begin{eqnarray}
\partial_t G &=&  [  p \sum_{g=1}^{j}  (j-g) n_{g}  + (1-j p)  +   \frac{G}{t} ], 
\end{eqnarray}
from which one finds that the transition occurs at $p_c=1/j$.

\section{p$<$1/2}
It is useful to introduce the time scale $d\tau = dt/(1+t)$ ($\tau \sim log(t)$) in which the set of equations to solve reads:
\begin{eqnarray}
\label{tau}
 \partial_\tau N_{1} &=&  (N_0 + p N_1 + p N_2) \cr
 \partial_\tau N_{g} &=&  [(1-p) N_{g-1} + p N_{g+1} ] .
\end{eqnarray}
This is a linear and homogeneous set of equations, so that one expects the solutions to have a time dependence $e^{\beta \tau} \sim t^{\beta}$, where $\beta$ is an eigenvalue of the dynamics . In the case $p>1/2$, we have shown above that $\beta=1$ is a proper eigenvalue and found the eigenvector Eq.\ref{eigenvector}. In the following, we look for the solution  $\beta(p)$ that is reached  when $p<1/2$. Solving the whole spectrum of eigenvalues of the matrix dynamics is out of question. Instead, we introduce the ansatz $N_i = A_i t^{\beta}$ and look for the solutions $A_i$:
\begin{eqnarray}
\label{rec}
\beta A_1  &=&  (p A_1  + p A_2 )  \cr
\beta A_g   &=&  [(1-p) A_{g-1}  + A_{g+1} p ] ,
\end{eqnarray}
which can be solved by recurrence:
\begin{eqnarray}
A_2  &=&  \frac{  \beta - p }{p} A_1 \cr
A_3 &=& \frac{ \beta^2 - \beta p  - p + p^2  }{p^2} A_1 
...
\end{eqnarray}

\begin{figure}
\includegraphics[angle=-90,totalheight=2.0in]{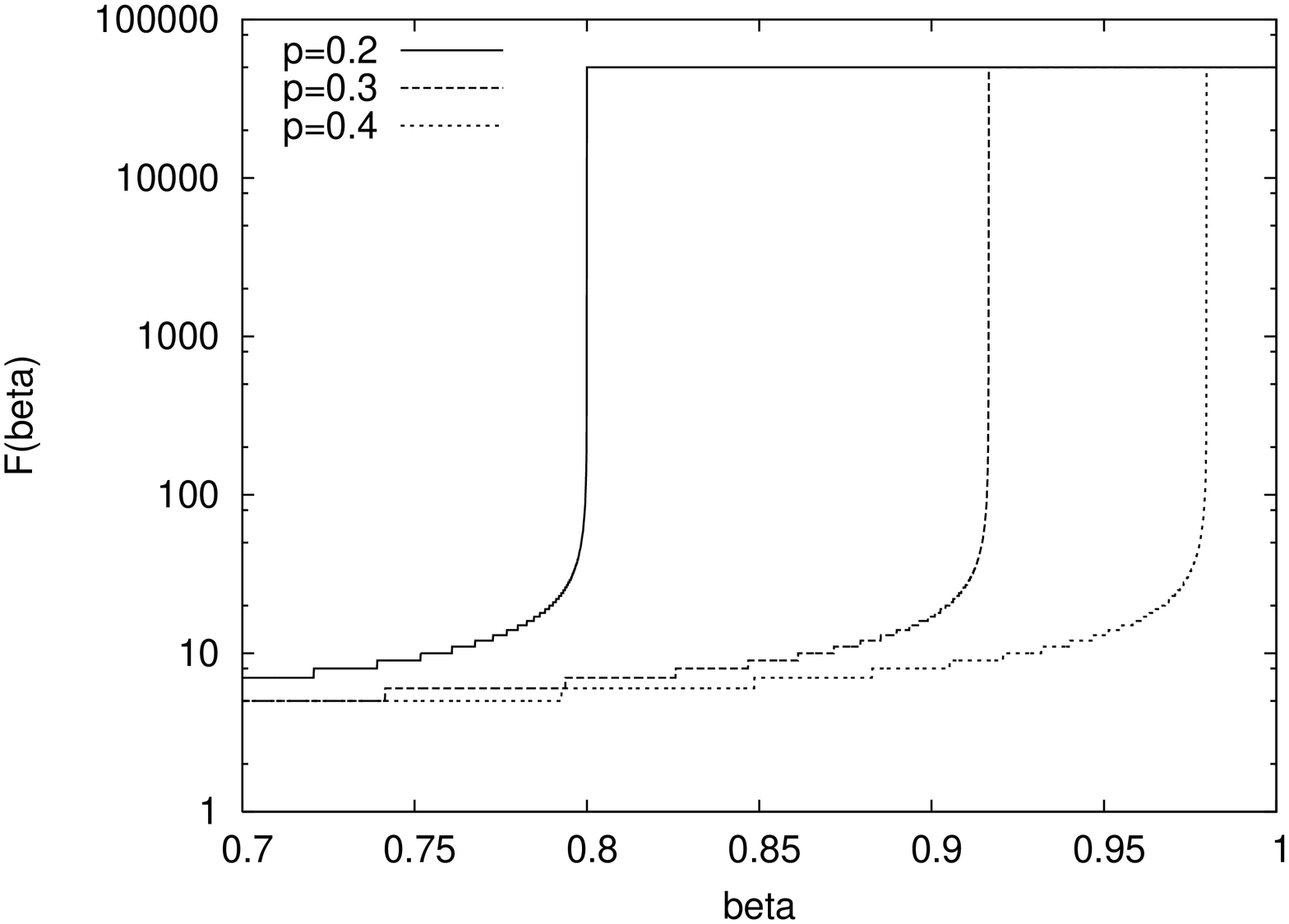}
\includegraphics[angle=-90,totalheight=2.0in]{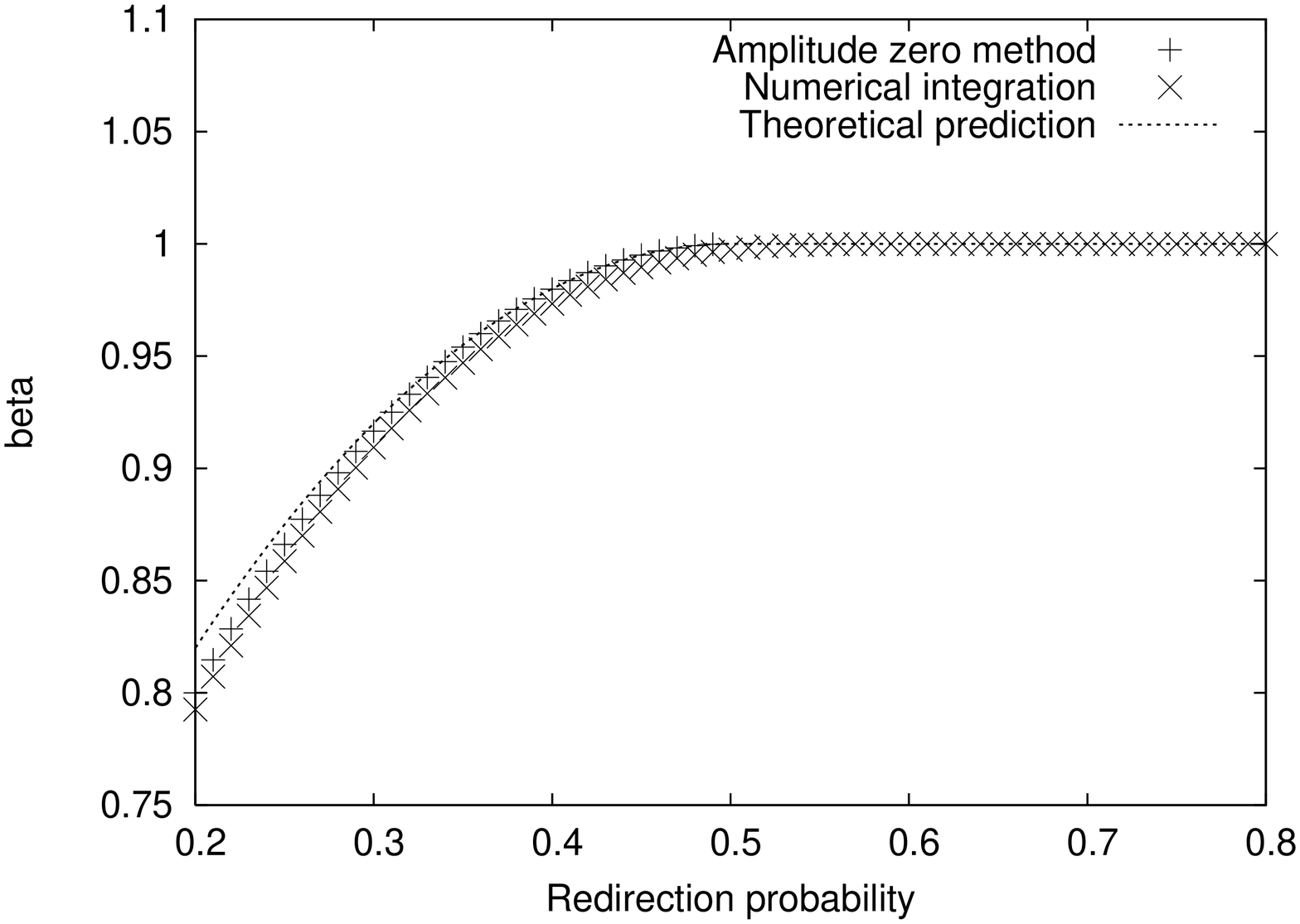}

\caption{In the left figure, relation between the index of the first negative amplitude $F(\beta)$ and the possible eigenvalue $\beta$. The results are obtained by numerically integrating Eq.\ref{rec} up to $g_{max}=5000$. The method shows that a whole region of $\beta$ exists where all $A_g$ are  positive. Let us stress that $F(\beta)$ is limited to the maximum value $g_{max}$ by construction. In the right figure, observed value of $\beta(p)$ obtained by integrating numerically the dynamical equations Eq.\ref{tau}. At $\tau=200$, the derivatives $D_i=\partial (\ln N_i)/\partial \tau$ are measured and are shown to be independent of $i$. Results are compared  with those obtained with the first negative amplitude approach. The small discrepancies are due to non-stationary effects, i.e. at $\tau=200$, the system has not yet reached its asymptotic state. The dotted line is the theoretical prediction Eq.\ref{tran}.}
\label{num}      
\end{figure}

A priori, any value of $\beta \in ]0,1[$ is available, except those for which any of the amplitudes $A_i$ becomes negative. In order to evaluate the values of $\beta$ that respect this condition,  we have  integrated numerically the above recurrence relations and looked, at a fixed value of $p$, for the relation $F(\beta)$, where $F$ is the index of the first amplitude $A_F$ that becomes negative, so that no amplitude $A_g$ is negative. By construction, $F(\beta)$  should go to infinity for allowed values of $\beta$. 
Numerical integrations (Fig.\ref{num}a) show that a whole region of $\beta < \beta_c$ are excluded due to this non-negativity constraint. In contrast, any value $\beta > \beta_c$ keeps all $A_i$ positive and is a priori susceptible to be chosen. However, numerical integration of Eq.\ref{rec} suggest that only this value $\beta_c$ is selected by the dynamics (Fig.\ref{num}b). In the limiting case $p\rightarrow 1/2$, the value $\beta=1$ is recovered.

Let us try to evaluate analytically the location of the transition. To do so, we focus on the relation
\begin{eqnarray}
\label{recLong}
\beta A_g   &=&  [(1-p) A_{g-1}  + A_{g+1} p ] 
\end{eqnarray}
for large values of $g$, assume that $A(g)$ is continuous and keep only the leading terms $A_{g+1} = A_g +  A^{'}_g + 1/2  A^{''}_g$. In this case, the recurrence relation recasts into the following homogeneous differential equation:
\begin{eqnarray}
\label{dif}
A^{''}_g   -  2 (1- 2 p) A^{'}_{g}  + 2 (1-\beta) A_g = 0.
\end{eqnarray}
It is straightforward to show that the solutions of this equation undergo a transition at:
\begin{eqnarray}
\label{tran}
\beta_c = \frac{1+4 p - 4 p^2}{2}.
\end{eqnarray}
Above this value, the amplitude $A_g$ is definite positive and asymptotically behaves like en exponential
$A_g \sim e^{(1- 2 p + \sqrt{\Delta})g }$,
where 
$\Delta=-1 +2 \beta - 4 p + 4 p^2$.
Below this value, in contrast, the solution exhibits an oscillatory behaviour
$A_g \sim e^{(1- 2 p) g}  e^{ i \gamma g }$, with $\gamma = \sqrt{1 -2 \beta + 4 p - 4 p^2}$. Consequently, this solution exhibits negative solutions, i.e. these values of $\beta$ are forbidden.
Comparison of this theoretical prediction with the numerical results (Fig.\ref{num}b) shows an excellent agreement, at least for small values of $1/2-p$. Let us stress, though, that deviations from this continuous approximation take place for large values of $1/2-p$. Indeed, $\beta_c$ goes to $1/2$ in the limit $p \rightarrow 0$, while one expects (and measures by integrating Eq.17) that  $\beta_c$ should go to zero.

\begin{figure}
\includegraphics[totalheight=2.0in]{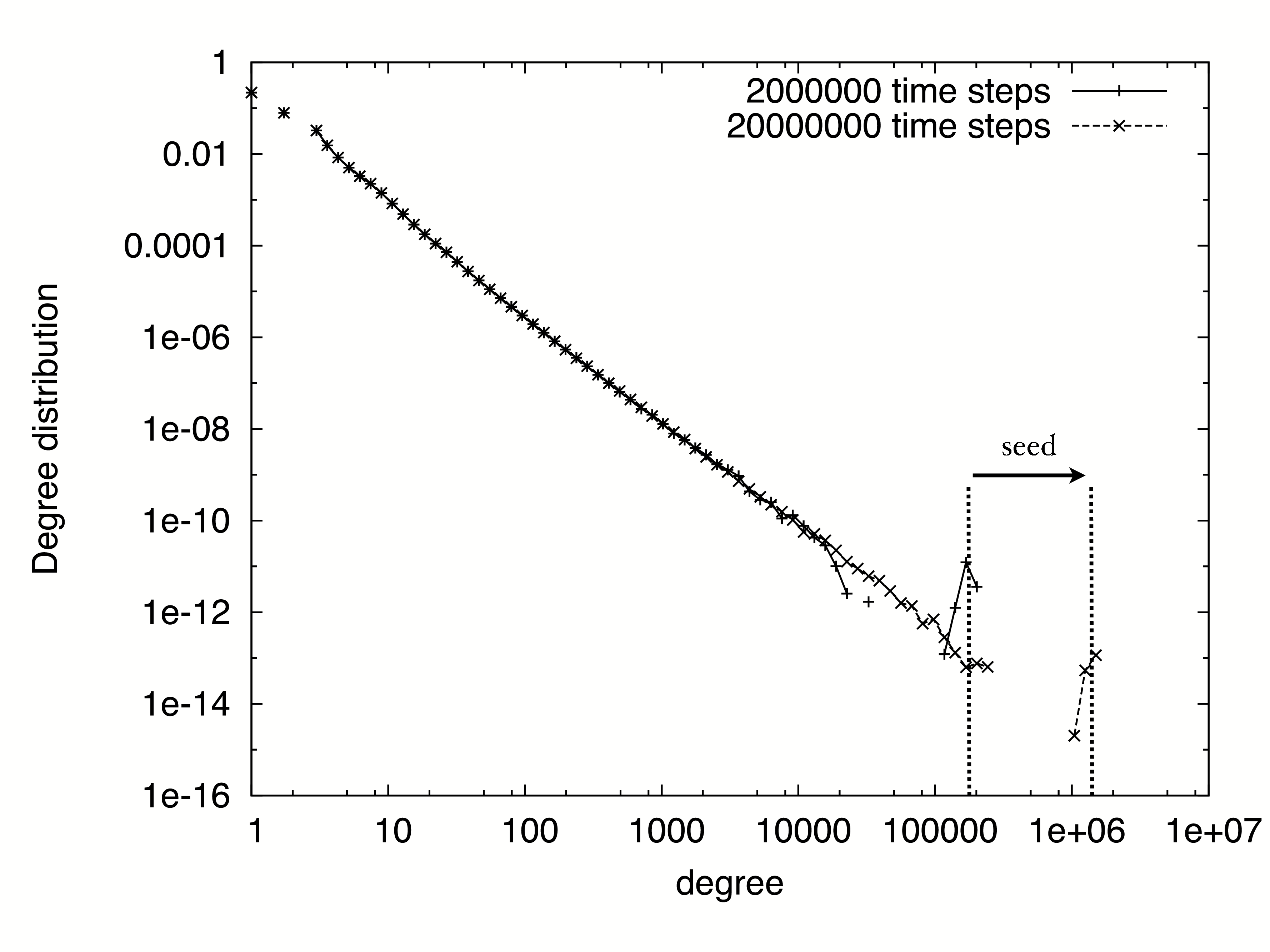}
\includegraphics[angle=-90,totalheight=2.0in]{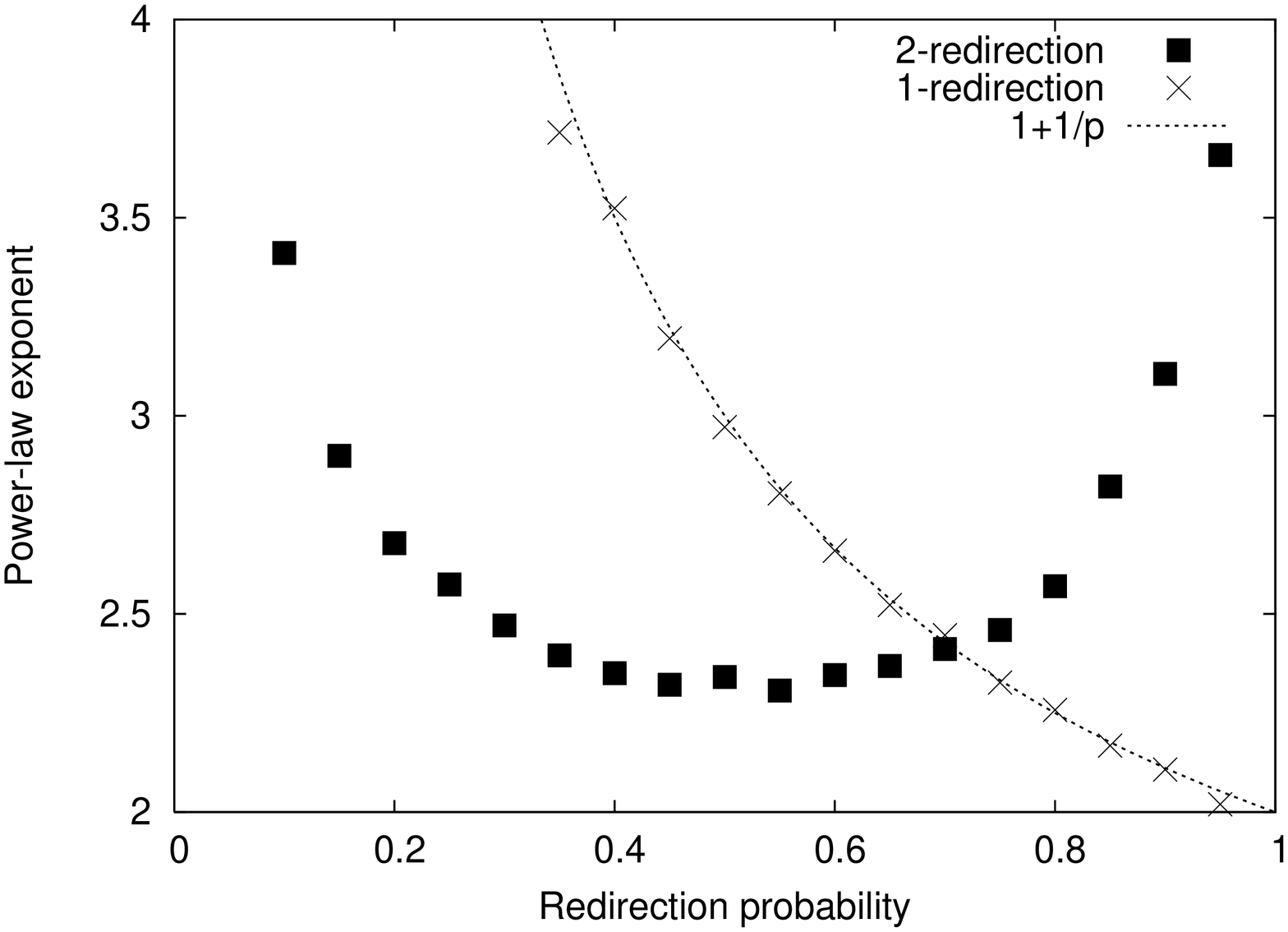}

\caption{In the left figure, degree distribution measured from simulations with $p=0.4$ at 2 different times $t=2 ~10^6$ and $t=2 ~10^7$. In the right figure, power-law exponent $\nu$ of the tail of the distribution $k^{-\nu}$.   The empirical result is compared with the theoretical result for the $1$-redirection model $1+1/p$.}
\label{num2}      
\end{figure}

\section{Degree distribution}

As soon as $j$-redirections with $j>1$ are introduced, the model exhibits complications in order to derive  a closed equation for the degree distribution. This is due to the fact that a {\em 2-variable} distribution for the degrees of the nodes at the extremities of one link  \cite{krapi} has to be added in order to account for the 2-redirections. Similarly, once one tries 
to write an equation for that distribution, the distribution involving three degrees characterizing two adjacent links has to be considered, etc., leading to an infinite hierarchy. A mean field description through a  truncature of the hierarchy at some level, even though possible in principle, has not been fulfilled yet and remains an open problem. In the following, we restrict the scope to a numerical analysis of the  degree distribution. To do so, we perform 50 computer realizations  of the random process,  measure the degree distribution after long times $t > 10^6$ and average over the many realizations.

Computer simulations show (Fig.\ref{num2}) that the distribution reaches  a stationary distribution except for a peak in its tail that advances in time. One observes that this peak velocity is $\sim t^{\beta_c}$, with $\beta_c(p)$ defined above. This result is expected as the average seed degree is $N_1$ and that this quantity grows like $t^{\beta_c}$.  Moreover, the stationary part of the distribution converges toward a power-law $k^{-\nu}$ for large values of $k$. We have verified the stationarity of this asymptotic state by measuring the degree distributions at different times $t$.

\section{Conclusion}
In this Letter, we have focused on a simple model of growing directed networks, where the probability for a node to receive a link depends on the number of paths of length $j$ arriving at this node. This process, that we called $j$-redirection, generalizes a redirection process  known to lead to preferential attachment \cite{krapi} and mimics the way people explore the Web. We have shown that when $j\geq2$, the system undergoes a transition to a regime where  condensates develop around the seed node. Condensates are nodes that receive a non-vanishing fraction of the links when the number of nodes $N$ goes to infinity. Let us stress that such states have been observed in other types of model \cite{con1,con2,con3,con4}, and that such {\em winner-takes-all} phenomena are associated to {\em extreme} configurations of the network, where a monopoly-like configuration develops. We have also focused on the degree distribution arising in such systems. Computer simulations show that the degree distribution asymptotically reaches an almost stationary state, where only the degree of the seed makes the solution unstationary. The stationary part is shown to converge to a power-law distribution. 
It is remarkable to note that the exponents belong to the interval $[2,3]$ for most of the values of the redirecting probability $p$. Let us stress that this effect reminds the properties of another model with redirection \cite{GNC}. The mechanism that we propose could therefore give an explanation for the proliferation of exponents in that interval \cite{ne,schnegg} in many empirical studies, e.g.  collaboration networks \cite{newman}, the Web \cite{comp}... To conclude, we would like to insist on the generality and simplicity of our approach, that is shown to exhibit a complex phenomenology.  As a next step, analytical predictions for the degree distributions, based on mean field assumptions, should be considered in order to improve our knowledge of the model.

 {\bf Acknowledgements}
R.L. has been supported by European Commission Project 
CREEN FP6-2003-NEST-Path-012864. R.L. would like to thank P. Krapivsky for fruitful (email) conversations.

\end{document}